\documentclass[letterpaper,10pt]{article}
\pdfoutput=1
\usepackage[italian,english]{babel}
\usepackage{amsmath}
\usepackage{amssymb}
\usepackage{braket}

\usepackage[a4paper,margin=2cm]{geometry}
\usepackage{graphicx}
\usepackage[leftcaption]{sidecap}
\usepackage{titling}
\usepackage{caption}

\usepackage{multicol}
\usepackage{tabularx}
\usepackage{booktabs}
\usepackage{xcolor}
\usepackage{dcolumn}
\usepackage[font=small]{caption}
\usepackage{subfig}
\usepackage{sectsty}
\sectionfont{\fontsize{12}{15}\selectfont}
\subsectionfont{\fontsize{11}{15}\selectfont}
\usepackage{tablefootnote}
\usepackage{float}

\usepackage{hyperref}\hypersetup{
colorlinks = true, 
urlcolor = blue, 
linkcolor = black, 
citecolor = blue 
}

\begin{document}

\selectlanguage{english}
\thanksmarkseries{arabic}

\title{ \textbf{\Large Visible pump - mid infrared pump - broadband probe: development and
characterization of a three-pulse setup for single-shot ultrafast
spectroscopy at 50 kHz}}

\author{\normalsize Angela Montanaro$^{1,2}$, Francesca Giusti$^{1,2,*}$, Matija Colja$^{2,3}$,\\ \normalsize Gabriele Brajnik$^{2}$, Alexandre M.A. Marciniak$^{1,2}$, Rudi Sergo$^{2}$,\\ \normalsize 
Dario De Angelis$^{2}$, Filippo Glerean$^{1,2}$, Giorgia Sparapassi$^{1,2}$, Giacomo Jarc$^{1,2}$, \\ \normalsize Sergio Carrato$^{3}$, Giuseppe Cautero$^{2,*}$, and Daniele Fausti$^{1,2,4,*}$\\
\\
\small \emph{${}^1$Department of Physics, Università degli Studi di Trieste, 34127 Trieste, Italy}\\
\small \emph{${}^2$Elettra Sincrotrone Trieste S.C.p.A., 34127 Basovizza, Trieste, Italy}\\
\small\emph{ ${}^4$Department of Chemistry, Princeton University, Princeton, New Jersey 08544, USA}\\
\small ${}^*$ francesca.giusti@elettra.eu, giuseppe.cautero@elettra.eu,daniele.fausti@elettra.eu \\
}

\date{June 5, 2020}
\maketitle

\pagenumbering{arabic}

\renewcommand{\abstractname}{} 
\begin{abstract}
\normalsize
We report here an experimental setup to perform three-pulse pump-probe measurements over a wide wavelength and temperature range. By combining two pump pulses in the visible (650-900 nm) and mid-IR (5-20 $\mu$m) range, with a broadband supercontinuum white-light probe, our apparatus enables both the combined selective excitation of different material degrees of freedom and a full time-dependent reconstruction of the non-equilibrium dielectric function of the sample. We describe here the optical setup, the cryogenic sample environment and the custom-made acquisition electronics capable of referenced single-pulse detection of broadband spectra at the maximum repetition rate of 50 kHz, achieving a sensitivity of the order of $10^{-4}$ over an integration time of 1 s. We demonstrate the performance of the setup by reporting data on mid-IR pump, optical push and broadband probe in a single-crystal of Bi$_2$Sr$_2$Y$_{0.08}$Ca$_{0.92}$Cu$_2$O$_{8+\delta}$ across the superconducting and pseudogap phases.
\end{abstract}

\begin{multicols}{2}

\section{Introduction}

Time-resolved optical techniques are a well-established tool to study a wide variety of ultrafast non-equilibrium dynamics in atoms, molecules and solids. In a standard pump-probe experiment, the sample under exam is photoexcited by an intense ultrashort pulse (pump) and the subsequent dynamical response is measured by a second, much weaker pulse (probe), properly delayed in time. As the cross-correlation of the employed pulses ultimately sets the temporal resolution of the experiment \cite{vardeny1981picosecond}, considerable effort has been put in the last decades in developing sub-picosecond sources, whose implementation enabled the investigation of extraordinarily fast processes, such as electron relaxation, charge transfer dynamics and vibrational coherence \cite{shah2013ultrafast,zewail2000femtochemistry,cabanillas2011pump,giannetti2016ultrafast}. Furthermore, the increasing spread of ultrafast optical parametric amplifiers (OPAs), and solutions based on non-linear optical processes in general, enabled a high frequency tunability of the pulsed laser source, so that optical properties of matter can be now probed from the THz to the XUV spectral range, even at high repetition rates \cite{cerullo2003ultrafast,bressler2004ultrafast,brida2009few,smith2011terahertz}. 

The possibility of properly tuning the energy of the photoexcitation opened up new, exciting applications in the field. If the bandwidth of the exciting pulse can be chosen to be on resonance with specific properties of the sample under study, different degrees of freedom can be selectively perturbed and monitored. This has a tremendous impact on femtochemistry (where source tunability allows to specifically target electron transitions and/or molecular vibrations, for instance), but also in condensed matter physics, where there is ample evidence that changing the pump frequency can lead to unexpected phenomena, for example in unconventional superconductors \cite{giannetti2016ultrafast,fausti2011light,mitrano2016possible,giusti2019signatures}.

The photon energy of the probe is also a crucial parameter of the experiment. In order to gain as much information as possible, a broad tunability of the probe is desirable to investigate relevant optical transitions occurring at different wavelengths. While the most intuitive approach would consist in acquiring pump-probe time-traces using a quasi-monochromatic probe beam and systematically changing the probing photon energy, this protocol requires long measuring times and potentially leads to altered measuring conditions, both in terms of laser system stability and sample environment. This has motivated the employment of broadband probe pulses that can encode the sample response over a wide wavelength window, and that can be generated either directly from dedicated OPA systems \cite{polli2007high}, via photonic crystal fibres \cite{dudley2006supercontinuum,cilento2010ultrafast} or self-phase modulation in appropriate transparent crystals \cite{ernsting2001wave,giannetti2009disentangling,novelli2012ultrafast,randi2016phase,novelli2017localized}. From an electronic point of view, this has stimulated interest in engineering silicon-based multichannel detection systems providing at the same time a sufficiently high number of pixels to ensure an adequate spectral resolution, and high frame rates to keep up with the increasing delivering rate of current pulsed sources. In this respect, advances in complementary metal oxide semiconductor (CMOS) and charge-coupled device (CCD) technology, enabling image sensors whose pixel readout can be as fast as 50 MHz, boosted the implementation of pump-probe schemes that can perform single-pulse acquisition at high repetition rates \cite{dobryakov2010femtosecond, aubock2012femtosecond,kanal2014100,baldini2016versatile}. 

In spite of the impressive variety of phenomena that can be investigated by simply adjusting the photon energy of the pulses involved, a fundamental limitation of standard pump-probe experiments is that they just allow to witness the dynamics of the energy redistribution triggered by the photoexcitation, without the possibility of actively controlling it. This liability inherently lies in the measuring procedure itself: the pump pulse that injects excitations in the sample and is meant to drive the system to off-equilibrium conditions (not adiabatically reachable otherwise), is the same pulse involved in the pump-probe scheme used to measure the effects of the perturbation. The pump-probe response is then embedded with both the order parameter dynamics and the energy relaxation processes, and these two contributions cannot be disentangled. In order to directly exert control and transiently shape the optical properties of matter, it is then crucial to decouple the photoexcitation from the measuring procedure. This has been attempted using a multi-pulse technique to study the emergence of superconducting and charge-density-wave order \cite{yusupov2010coherent,madan2014separating,madan2016real}, by using a combination of three quasi-monochromatic pulses, all centered at 800 nm: a first, very intense pulse initially destroys the ordered state and then the subsequent dynamical evolution is monitored by a weaker pump-probe sequence. 

In this work, we present a versatile setup to implement a three-pulse scheme to perform non-equilibrium broadband optical spectroscopy. The unique feature of our technique consists in combining, in the very same measurement, two pumping pulses having tunable photon energy in two different regions of the electromagnetic spectrum: the first beam can be tuned in the visible range (650-900 nm) to be resonant with the electronic degrees of freedom of the material, while the second one spans the mid-infrared region (5-20 $\mu$m) thus covering low-energy excitations. The visible and mid-IR responses can be either singled out or jointly analysed, allowing for a direct and selective control over different degrees of freedom of the sample and their possible coupling. The system is probed by a supercontinuum white-light (500-800 nm) whose referenced single-pulse detection is enabled by a novel, custom-made acquisition system based on an analog-to-digital converter (ADC) and a field-programmable gate array (FPGA) acquiring up to 50000 spectra/s. The setup has been optimized to perform systematic transient reflectivity or transmittivity studies on crystalline samples over a wide range of temperatures (33-300 K) and sample azimuthal orientations (0-360$^{\circ}$). 

We test the setup capabilities by performing transient reflectivity measurements on a slightly underdoped sample of Bi$_2$Sr$_2$Y$_{0.08}$Ca$_{0.92}$Cu$_2$O$_{8+\delta}$ (Y-Bi2212), showing that one single measurement is sufficient to study the wavelength-dependent dynamics of the coupling between low- and high-energy excitations across the superconducting and the pseudogap phases.

\section{\label{sec:Exp_setup}Experimental setup}
\subsection{\label{sec:Laser_sys}Laser system and optical arrangement}
\begin{figure*}
    \centering
    \includegraphics[scale=0.5]{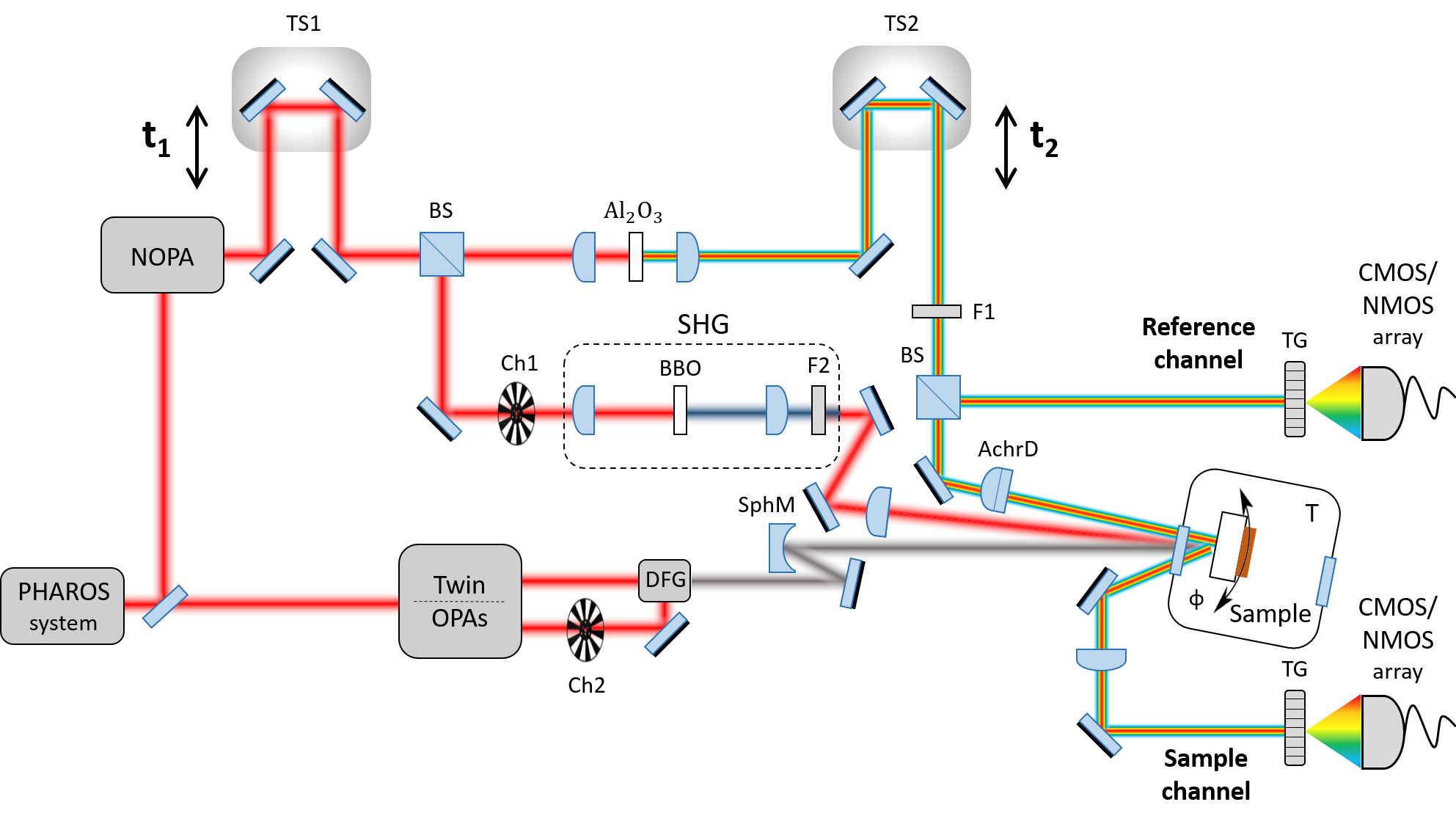}
    \caption{Schematic representation of the experimental setup. A detailed description of the optical arrangement and the labels assigned to the optical elements is in the text.}
    \label{fig:setup}
\end{figure*}

A block diagram of the optical setup is shown in Figure \ref{fig:setup}. The laser source consists of a Non-Collinear Optical Parametric Amplifier (Orpheus-N by Light Conversion) and a Twin Optical Parametric Amplifier (Orpheus TWIN by Light Conversion). Both systems are pumped by the Pharos Laser (Light Conversion) which delivers 400 $\mu$J pulses with 1.2 eV photon energy (290 fs). The NOPA is pumped with 40 $\mu$J/pp while the remaining 360 $\mu$J/pp pump the Twin OPAs. The repetition rate of the laser source is tunable from 50 kHz down to single shot using a built-in pulse-picker. 

The NOPA output is tunable from a minimum of 650 nm to a maximum of 900 nm and delivers pulses shorter than 25 fs on the entire range. The carrier envelope phase stable mid-IR pump is obtained through Difference Frequency Generation in a GaSe crystal by mixing the outputs of the Twin OPAs and is tunable from 5 to 20 $\mu$m.

The NOPA output is split by a 70/30 beam splitter: most of the power is used to pump the sample and the remaining part to generate the white-light probe. The setup has been designed to work in two main configurations: the user can be either choose to pump the system directly using the light pulses exiting the NOPA or exploit Second Harmonic Generation (SHG) to achieve higher photon energies.  In the latter case, the pump beam is obtained by focusing the NOPA beam in a 0.7 mm thick type I $\beta$-Barium Borate (BBO) crystal, optimized for SHG with 900 nm. After collimation, a bandpass coloured filter (F2, 350-700 nm) filters out the remaining near-IR radiation and the blue beam is then focused on the sample surface. The white-light probe is generated via Self Phase Modulation by focusing (5 cm focal length) the beam in a 2 mm thick sapphire (Al$_2$O$_3$) crystal. A stable white light regime is reached by tuning an iris placed in front of the focusing lens. A shortpass filter (F1, edge at 800 nm) filters out the residual component of the generating near-IR light. The final spectrum of the supercontinuum white-light ranges from about 500 to 800 nm.  In order to improve the white-light pulse-to-pulse stability, we have introduced in the detection scheme a reference channel. Before impinging the sample, the probe beam is split, and the reflected part (a pristine copy of the probe beam) is routed towards a twin detector to be simultaneously acquired. The transmitted probe beam is instead focused on the sample using an achromatic doublet to avoid chromatic aberrations. The mid-IR pump is obtained by filtering the DFG output through a germanium window which absorbs photons with energy above 0.3 eV. The beam is then focused on the sample by means of a silver spherical mirror.
The focus size of the three beams at the sample surface has been measured by knife-edge method (5 $\mu$m accuracy). Typical values of the full width at half maximum of the three beam profiles are 100 $\mu$m for the visible pump, 150 $\mu$m for the mid-IR pump and 70 $\mu$m for the probe. We estimated the visible pump temporal length to be $\sim$35 fs via SHG-FROG. When SHG configuration is used, the visible pump duration slightly increases to reach $\sim$50 fs (measured by cross-correlation measurements). The mid-IR pump time duration is wavelength-dependent and shorter than 150 fs on all the energy range. By default, the mid-IR pump is vertically polarized. The polarizations and the intensity of the probe and the visible pump can be independently adjusted using two pairs of half-wave retardation plates and polarizers. The setup has been designed to work in a non-collinear geometry: the probe beam impinges almost normally to the sample surface; the visible pump propagation direction makes an angle of about 10$^{\circ}$ with the probe and the mid-IR beam impinges onto the sample surface at an angle of 20$^{\circ}$. 

The setup is provided with two translation stages to independently control the time delay between the three beams. A first translation stage (TS1) is placed at the NOPA output, before the first beam splitter, so that TS1 tunes the delay between the two pumps, without affecting the delay between the probe and the visible one. The latter is controlled by a second translation stage (TS2) which modifies the optical path of the white light probe only. In a standard experiment, for each given position of TS1, a full scan of TS2 is performed.
Finally, we report in Table \ref{tab:setup_performance} the standard optical parameters which summarize the performance of the  setup.

\begin{table*}
\caption{\label{tab:setup_performance}Typical performance of the setup.}
\begin{tabularx}{\textwidth}{ccccc}
\hline\hline
 Pulse&Tunability&Energy per pulse&Time duration
&Polarization\\ \hline
 Visible pump & 650-900 nm & $<$1.4 $\mu$J & 35 fs & Tunable \\
 & (or SHG configuration available) & $<$200 nJ (SHG) & & \\
 Mid-IR pump& 5-20 $\mu$m & 160 nJ (@17 $\mu$m) & 150 fs & Vertical \\
 & & 1.4 $\mu$J (@9 $\mu$m) & & \\
 Broadband probe& 500-800 nm & $<$6 nJ & 1.5 ps$^a$
 &Tunable \\
 & (generated @860 nm) & & (chirped) & \\
 \hline\hline
\end{tabularx}
\footnotesize{$^a$ It is worth highlighting that the relatively long time duration of the chirped white light pulse does not compromise the overall temporal resolution of the experiment, as demonstrated in ref. \cite{polli2010effective}.}
\end{table*}

\subsection{\label{sec:Sample}Sample environment}
The sample is mounted in a closed cycle liquid helium cryostat (DE 204 by Advanced Research Systems). The cryostat expander is supported by a custom-made structure fixed to the laboratory floor; the cold head is instead fixed to the optical bench. This design allows to efficiently isolate the sample from the strong vibrations experienced by the expander. The custom-made cryostat structure is compatible with adjustments along the x, y and z directions.

The sample holder consists of a copper plate which is directly connected to the cold head. On top of this plate, we fixed a piezo-electric rotator (Attocube, ANR240) which allows a rotation of the sample by the azimuthal angle $\phi$. Samples are finally mounted on a small copper disk – fixed to the moving part of the rotator - using fast-drying silver paint. We show a picture of the final sample’s arrangement in Figure \ref{fig:sample}.  A thermocouple, fixed in the sample proximity, is used to
\begin{center}
    \includegraphics[scale=0.4]{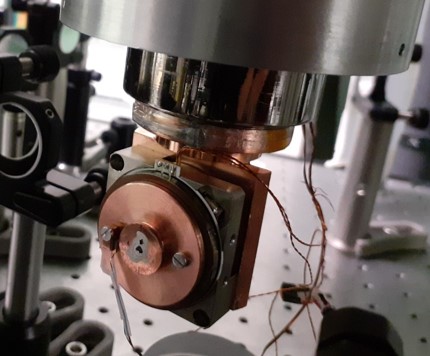}
    \captionof{figure}{Single-crystal Y-Bi2212 samples mounted on the piezo-electric rotator. A thermocouple is fixed in close proximity of the sample by a small copper clamp.}
    \label{fig:sample}
\end{center}
directly measure the sample temperature. Indium foils have been interposed at each metallic interface to improve the overall thermal conductivity.

The structure in Figure \ref{fig:sample} is enclosed in a vacuum chamber that allows optical access through a front 1 inch window. The composition of the window is chosen according to the photon energies of the three beams involved. Because of the presence of the mid-infrared pump, we have chosen to mount a polycrystalline diamond window, which is 0.5 mm thick. The vacuum chamber is provided with other identical apertures, in correspondence of each chamber’s face. This provides the possibility of performing transmittivity measurements, if required. In this respect, it is worth mentioning that the piezo-electric rotator has a hole in the middle so that transmittivity measurements can be performed without compromising the azimuthal degree of freedom.

Vacuum conditions are matched via a standard turbo pumping station (Pfeiffer, HiCube 80 Eco). Pressures as low as 10$^{-7}$ mbar can be reached when working at ambient temperature; at cryogenic temperature, the usual working pressure is 10$^{-8}$ mbar. While the minimum attainable temperature is 12 K when the sample is directly fixed on the first copper plate (so very close to the cold head), it increases up to 33 K when using the piezo-electric rotator. A complete cooling cycle takes about 4 hours. A temperature controller provided with a feedback circuit enables the user to remotely modify the temperature of the sample, so that a full temperature scan, i.e. the acquisition of a simple pump-probe trace for each temperature within a given
 \begin{center}
    \centering
    \includegraphics[scale=0.45]{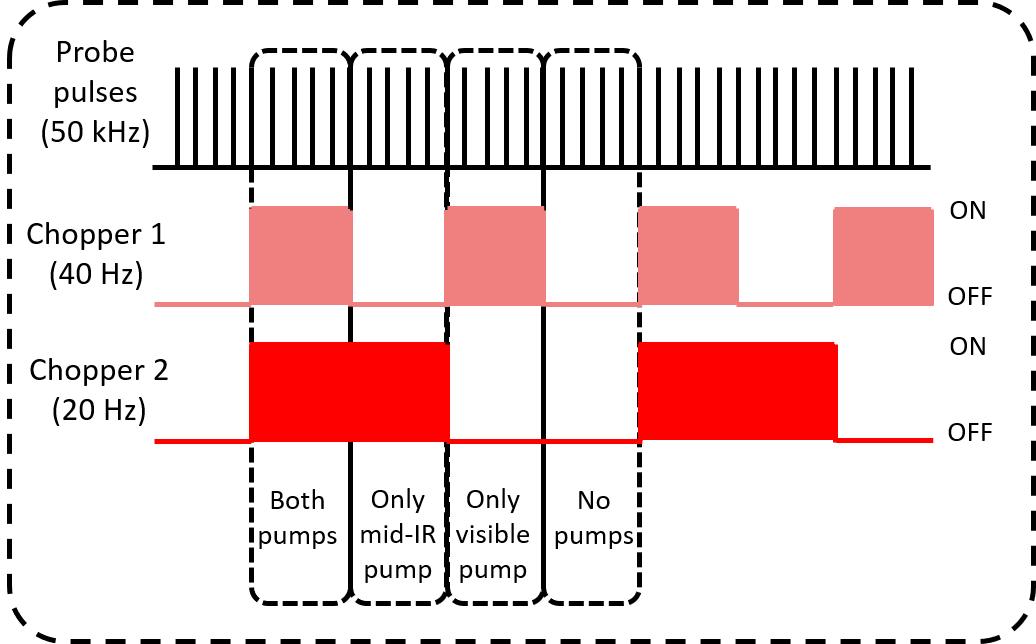}
    \captionof{figure}{A simple sketch of the choppers’ operation designed to reference the acquisition in different pumping conditions.}
    \label{fig:chopper}
\end{center}
range, can be performed.

\subsection{\label{sec:detection}Detection and chopping scheme}
 The reflected (or transmitted) probe beam is collimated by a lens and routed towards the acquisition part, designed for a frequency-resolved, pulse-by-pulse detection. The beam is dispersed by a transmission blazed grating (TG, in Figure \ref{fig:setup}) and then focused on the detector. According to the user’s needs, two options of detectors are available: a complementary metal oxide semiconductor (CMOS) linear image sensor (Hamamatsu S11105 \cite{CMOS}) made up of 512 pixels, or a negative channel metal oxide semiconductor (NMOS) linear image sensor (Hamamatsu S8380-128Q \cite{NMOS}) provided with 128 pixels. The detector acquisition is synchronized with the laser repetition rate and can run up to 50 kHz (when CMOS are needed) or up to 5 kHz (when NMOS are installed). A detailed description of the acquisition system and electronics required to achieve this performance is given in Section \ref{sec:Acquisition}. A new wavelength calibration of the arrays is performed every time the optical alignment is adjusted.  To do so, a common coloured filter having distinctive transmittivity features across the white light bandwidth can be used (FGB67 by Thorlabs, for instance). The filter is usually placed after the collimating lens behind the sample; a spectrum averaged over thousands of pulses is acquired using first the array photodiode detector, and after an already calibrated fiber spectrometer. The final calibration is then achieved by interpolation of the two spectra.
  \begin{center}
    \centering
    \includegraphics[scale=0.45]{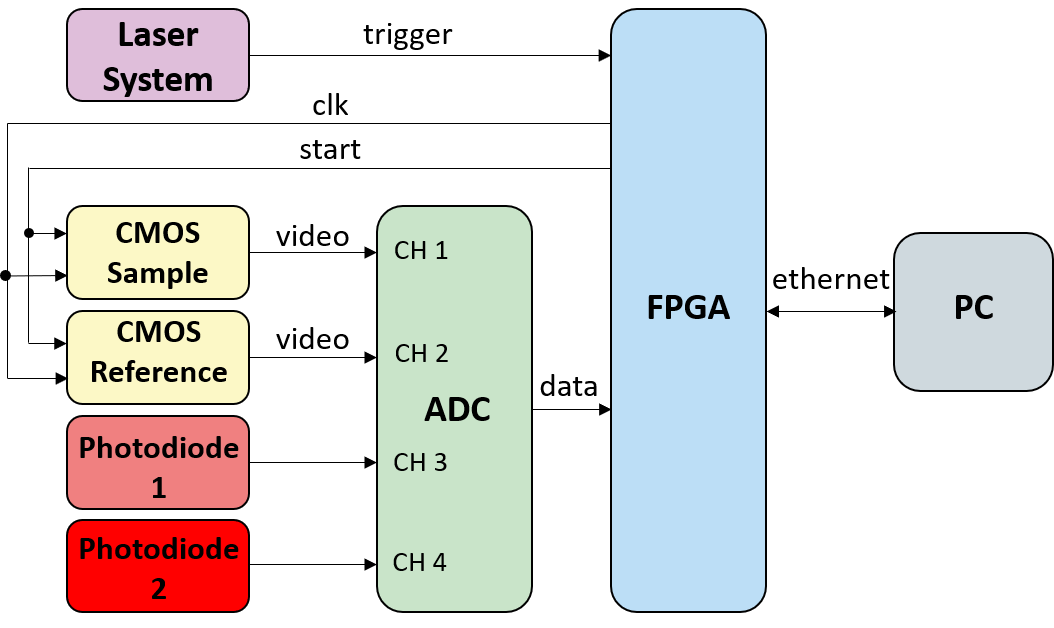}
    \captionof{figure}{Schematic representation of the acquisition electronics synchronized with the laser repetition rate.}
    \label{fig:electronics}
\end{center}
To improve the detection of the signal, two choppers have been inserted along the optical paths of the two pumps. The first chopper (Ch1 in Figure \ref{fig:setup}), periodically blocking the visible pump propagation, runs at 40 Hz, while the second one (Ch2 in Figure \ref{fig:setup}) runs at 20 Hz. The choppers’ controllers are both referenced with the same square wave generator, so that the mechanical rotations of the blades are automatically synchronized. A simplified sketch of the chopping scheme is shown in Figure \ref{fig:chopper}. Each single probe pulse recorded by the photodiode arrays (sample and reference channels) can fall into one of the four dashed boxes drawn in Figure \ref{fig:chopper}. In order to sort the pulses according to this scheme, we use two single-photodiode detectors (PhD1 and PhD2). The photodiodes are properly placed to intercept a back-reflection or a residual transmission of the two pump beams, and their signals are digitized by the ADC (CH2 and CH3 in Figure 3). These signals give a feedback about the chopper status (ON/OFF) and ultimately provide a tool to assign each probe pulse detected by the array detectors with a label, indicating which pump beam has interacted with the sample prior to its arrival.
 
 \section{\label{sec:Acquisition}Acquisition system }
As highlighted above, the setup is provided with two pairs of photodiode arrays, based on CMOS or NMOS linear image sensors. The main difference between the two detection systems lies in the maximum video data rate (and thus the single-shot repetition rate) that can be achieved: while the NMOS data rate is limited to 2 MHz, the CMOS arrays belong to the high-speed Hamamatsu sensor family with video data rate as high as 50 MHz. We will limit here our discussion to the CMOS arrays only, as their acquisition at 50 kHz is the most challenging to handle from an electronic point of view. The same scheme can be adapted to the NMOS detection, with the caveat that an additional external amplifier (Hamamatsu C7884 series) is needed and the clock pulse frequency has to be adjusted.

\subsection{\label{sec:electronics}Electronics}
A dedicated electronic setup has been developed by the Detectors \& Instrumentation Laboratory of Elettra Sincrotrone Trieste. A schematic representation of the system is shown in Figure \ref{fig:electronics}.

The CMOS sensors are mounted on custom-made boards. Each of the 512 photodiodes is directly connected to a charge-sensitive-amplifier (CSA) with a 0.1 pF feedback capacitor, which determines a high conversion gain. The device also integrates buffer amplifiers, digital shift registers and a timing generator that allows operation only with start and clock inputs.

\begin{figure*}
    \centering
    \includegraphics[scale=2.5]{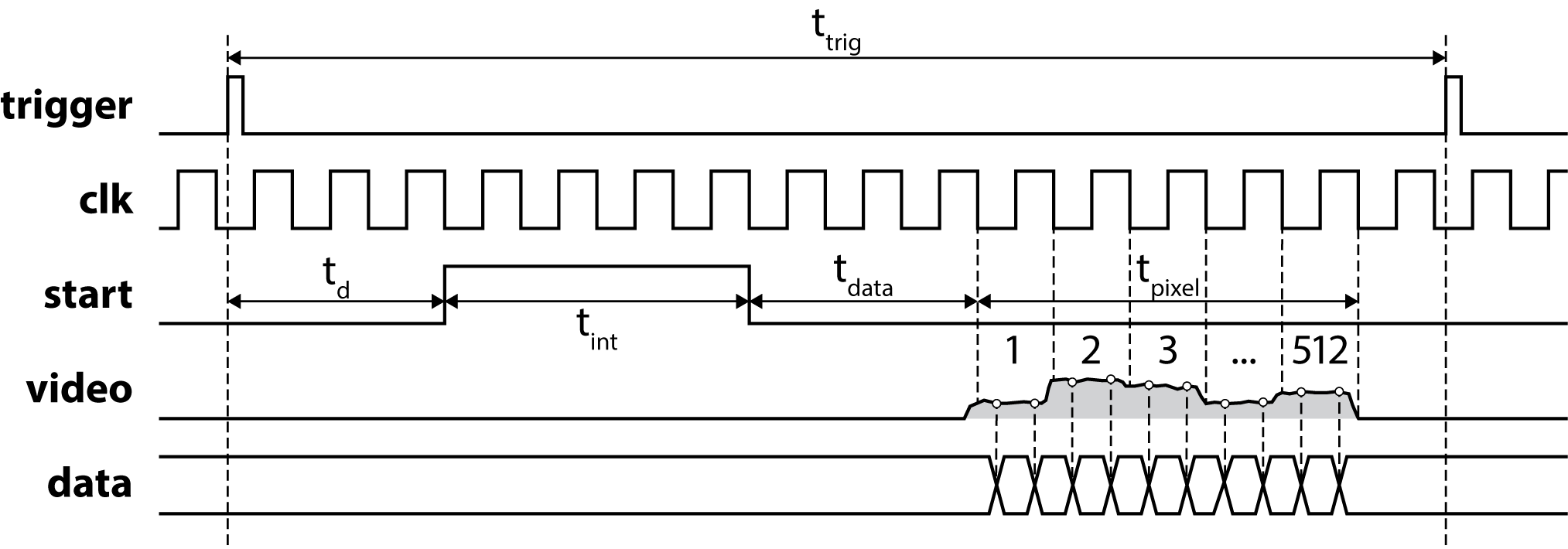}
    \caption{Timing diagram summarizing the operation of the CMOS linear image sensors.}
    \label{fig:timing}
\end{figure*}

The core part of the acquisition system is the FPGA (Arria 10 GX by Intel) which controls the CMOS sensor operation, coordinates the ADC acquisition and handles data transfer to the PC through an Ethernet link. More precisely, the FPGA continuously generates the 50 MHz clock needed by the sensor and waits for the trigger pulse periodically generated by the laser system (t$_{trig}$ = 20 $\mu$s, 50 kHz). The start pulse generation is delayed with respect to the trigger, and this delay (t$_d$) can be dynamically set in order to synchronize the acquisition with the probe pulse. A rising edge on the CMOS start pin sets the simultaneous integration of all pixels. At falling edge, the collected charge, converted into a voltage through the CSA amplifier, is transferred to the hold circuit and the internal shift register begins to sequentially provide data on video output. The start pulse high period (t$_{int}$) sets the integration time and determines the number of pixels to be read. In our case, we want to read all the pixels and, at the same time, keep the integration time as small as possible to collect only the light of the probe pulse. Considering these requirements, we set the integration time to 512 clock periods, that is 10.24 $\mu$s. Since the readout and integration circuits in the sensor are separated from each other, a new integration can be performed while data transfer is still in progress. Once the start pin has returned to zero the first pixel data is given on the next 49$^{th}$ clock falling edge (t$_{data}$ = 980 ns) and the readout operation continues for 512 clock periods (t$_{pixel}$ = 10.24 $\mu$s). This behaviour is summarized in the timing diagram shown in Figure \ref{fig:timing}.

\subsection{\label{sec:data_transm}Data acquisition and transmission}
In order to perform referenced measurements and to acquire the output of the two photodiodes (PhD1 and PhD2), we decided to utilize a 4-channel 16-bit analog-to-digital converter (AD9653 by Analog Devices) which operates at a conversion rate up to 125 MSPS with a 650 MHz analog bandwidth. However, the CMOS output voltage characteristics are not compatible with the input specifications of the ADC, so it cannot be connected directly to the ADC. In fact, CMOS output has a saturation voltage of about 1.3$\pm$0.6 V with an offset of 1.2$\pm$0.6 V, which could saturate the ADC and even damage it. For this reason, we designed a front-end circuitry to compensate the offset voltage of the video signal utilizing a digital-to-analog converter (AD5541A by Analog Devices).

The sampling frequency has been fixed to 100 MHz and it is generated by a dedicated low-jitter clock generator (SI5340 by Silicon Labs). Therefore, for each analog value provided by CMOS output we acquire 2 samples, thus obtaining 1024 samples. The data of each channel are then internally serialized inside the ADC and provided to the FPGA using LVDS differential pairs (Low Voltage Differential Signalling) in DDR mode. All these components are mounted on an internally developed FMC card (FPGA mezzanine card) which has been connected to an FMC carrier board, custom made as well, based on a high-performance Arria 10 GX FPGA, capable to host two FMC modules, with an 8 GB DDR3 RAM onboard and high-speed serial data links (1/10 Gb Ethernet).

An appropriate Verilog code takes care of every task: generation of CMOS timing signals, DAC programming, synchronous data acquisition from ADC at every trigger event, storage of captured events in RAM and transmission of collected data through a User Datagram Protocol (UDP) Ethernet link, are performed using a dedicated FPGA hardware. Even if the intrinsic parallel architecture of the FPGA allows to treat large amounts of data sets and their transmission, a great effort has to be done in order to correctly receive and store them. In fact, considering a stream of 1024 16-bit samples multiplied by four channels with a laser repetition rate of 50 kHz, the resulting data rate is 3.3 Gb/s (410 MB/s). A typical point-to-point 1 Gb/s Ethernet connection between FPGA and PC and a standard mechanical hard disk drive cannot handle such a high data rate, so we moved to 10 Gb/s Ethernet and SSD (Solid State Disk) drives.

\section{\label{sec:Meas}Measurements}
\subsection{\label{sec:Data_proc}Data structure and processing}
We underlined above that our setup is suitable for both reflection and transmission geometries, depending on specific requirements set out by the sample under exam (bulk crystal, thin films, solutions). Here, for simplicity, we restrict our discussion to transient reflectivity measurements only.

\begin{figure*}
    \centering
    \includegraphics[scale=0.4]{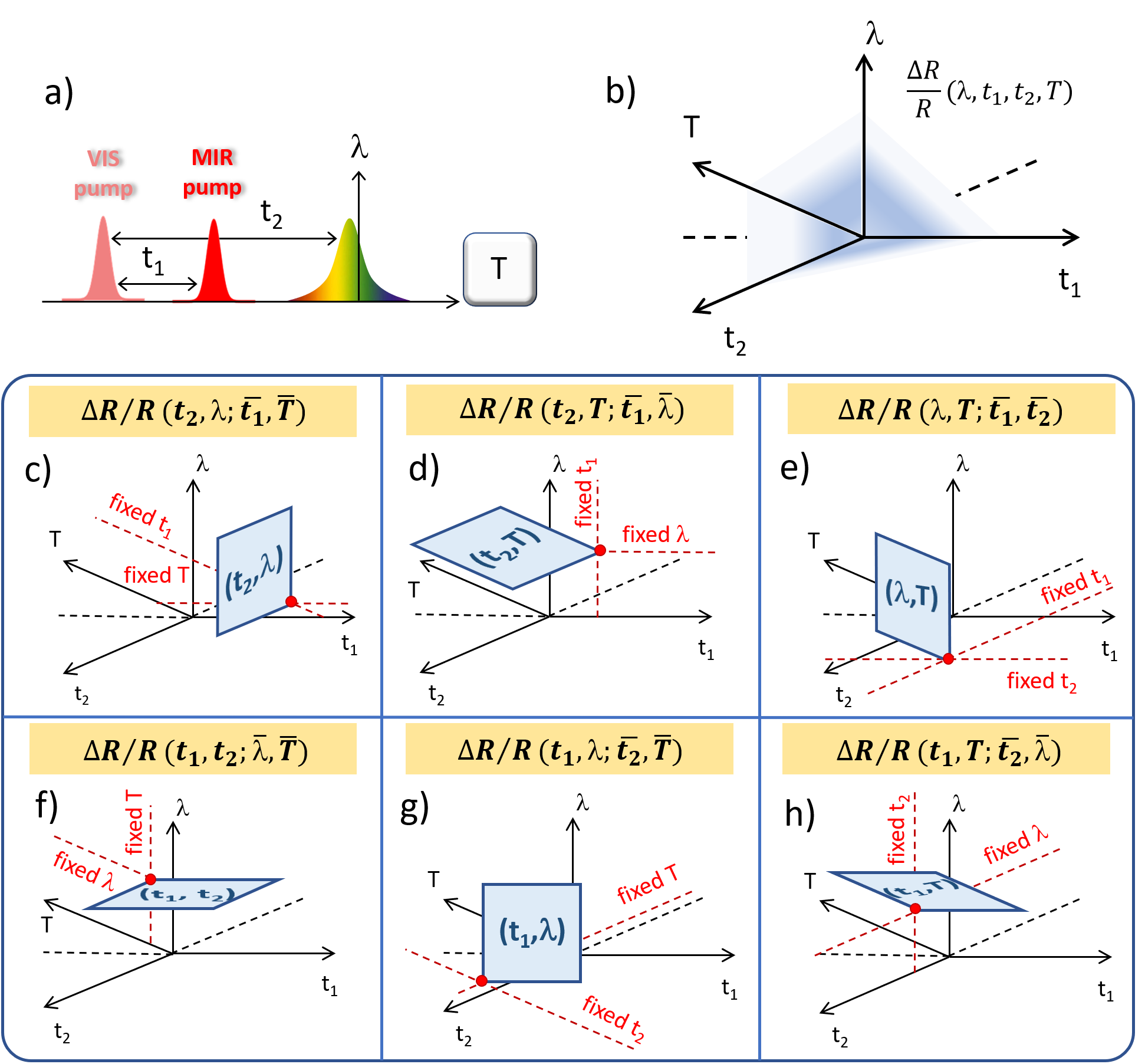}
    \caption{Details of the data structure. a) Depiction of the temporal arrival of the three pulses at the sample (at a given temperature T) and their relative time delays. We highlight that $t_1$ is $<$ 0 ( $>$ 0) when the mid-IR pulse interacts with the sample before (after) the arrival of the visible pump. b) Relevant degrees of freedom of the four-dimensional hypercube representing our dataset. c-h) Two-dimensional slices of the full dataset obtained by fixing two parameters at a time; there are six possible combinations. The transient reflectivity can be displayed as function of time delay and wavelength of the probe (c), time delay of the probe and temperature of the sample (d), probe wavelength and temperature (e),  relative delay between the two pumps and probe delay (f), probe wavelength (g) and sample temperature (h).}
    \label{fig:data_structure}
\end{figure*}

The measured quantity in our experiment is the $-i^{th}$ frequency-dispersed reflected probe pulse $R_i^s(\lambda,t_1,t_2,T)$, namely the intensity recorded by the linear array detector along the sample channel at each pixel, for a given delay $t_1$ between the two pumps, a given delay $t_2$ between the white-light probe and the visible pump, and a given sample temperature $T$ measured by the thermocouple. In Figure \ref{fig:data_structure}a, we provide a simple scheme which summarizes the degrees of freedom of the measurement and their physical meaning. We recall that we also simultaneously acquire a copy of the pristine probe pulse along the reference channel, $R_i^r(\lambda,t_1,t_2,T)$. Our optical observable is then the referenced average differential reflectivity change, defined as follows:

\footnotesize
\begin{equation}
\frac{\Delta R}{R}(\lambda, t_1, t_2, T) = \frac{\sum _{i=1} ^N R^{s,P}_i(\lambda, t_1, t_2, T)}{\sum _{i=1} ^N R^{r,P}_i(\lambda, t_1, t_2, T)} - \frac{\sum _{i=1} ^N R^{s,U}_i(\lambda, t_1, t_2, T)}{\sum _{i=1} ^N R^{r,U}_i(\lambda, t_1, t_2, T)} \label{eq:deltaR}
\end{equation}
\normalsize
where $N$ is the number of pulses over which the integration is performed, and the superscripts $P,U$ refer to “pumped” and “unpumped” probe pulses, respectively. By “unpumped” we refer to those pulses reflected by the sample when both pumps are blocked by the choppers (pulses falling in the last box in Figure \ref{fig:chopper}). Furthermore, we report that each single curve in Equation \ref{eq:deltaR} is systematically corrected for a background spectrum R$_{BG}(\lambda)$ which is acquired before each measurement by blocking the probe beam and let the two pumps open. The background subtraction serves the purpose of cleaning the acquired spectra from any scattered light coming from the pumps and from the environment.

The structure of our data is therefore a four-dimensional hypercube, as depicted in Figure \ref{fig:data_structure}b, where all the four axes are orthogonal to each other. Given the big amount of data, the most straightforward way to visualize the outcome of measurement consists in dissecting the hypercube into two-dimensional slices, by fixing two parameters at a time. 

We show in Figure \ref{fig:data_structure}c-h all the six possible combinations that can be computed in the four-dimensional space.  We will refer to these schemes in the following, and use the notation $\Delta R/R(w,x;\bar{y},\bar{z})$, where $(w,x)$ are the variables whose dependency is studied and $(\bar{y},\bar{z})$ the ones that have been kept fixed.

Recalling that we perform a sorting of the probe pulses according to the chopping scheme in Figure \ref{fig:chopper}, there are three different typologies of “pumped” pulses, so that, for each single measurement, we can construct three different four-dimensional maps: i) a map which contains a two-colour signal only due to the effect of the visible pump (VIS-MAP); ii) a map displaying the two-colour signal due to the interaction of the sample only with the mid-IR (MIR-MAP); iii) a map showing the effect of applying both pumps (PP-MAP). For each of
the three types of maps, an average time-independent signal is calculated over the negative-t$_2$ region and systematically subtracted. It is worth mentioning that, even though all three typologies of maps are technically four-dimensional hypercubes, the relative delay between the two pumps (t$_1$) is a relevant degree of freedom only for the PP-MAP, which actually contains a three-colour signal. Consequently, the analysis of the VIS-MAP and the MIR-MAP will not require the computation of the three two-dimensional slices for which t1 is a variable parameter (i.e. combinations from f to g in Figure \ref{fig:data_structure}). 

Since no physical correction of the chirp of the broadband white light pulses was performed, each of these three maps has to be post-processed to compensate for the dispersion. After white-light generation, we estimate the probe beam’s temporal length to be about 1.5 ps at the sample surface. We stress that as we are frequency resolving the signal, the temporal resolution of the setup for a given probe frequency is significantly better than 20 fs \cite{polli2010effective}. The correction is performed numerically by shifting each line in the four-dimensional dataset $\Delta R/R(\lambda,t_1,t_2,T)$ along the t$_2$-axis. The amount of the shift is quantified by defining a zero-time-delay vector (one value for each wavelength) by looking at the VIS-MAP. We use the onset of the two-colour signal in the VIS-MAP as a reference point for the temporal overlap between the visible pump and the broadband probe\footnote{We highlight that additional care must be eventually taken when measuring a transmittivity signal, since a further correction must be performed to compensate for the velocity mismatch inside the sample.}. 

We recall that the novelty of our setup consists in being able to uncover possible modifications in the dynamical signal due to the combined action of the visible and the mid-IR pumps. In order to access this information, we perform a point-to-point subtraction between the PP-MAP (which contains the effects of both of them), and the VIS-MAP and MIR-MAP (which contain the effects of the two pumps when they act independently on the sample).  We will refer to this new four-dimensional map as “differential map” (DIFF-MAP) in the following.

Finally, our setup is characterized by an intrinsic noise level of $10^{-4}$ rms over an integration time of 1 s. This signal-to-noise ratio can be further improved statistically by either increasing the number of shots to be averaged over or repeating each experiment many times. Both choices come obviously at the cost of significantly reducing the speed of the measurement. The robust experimental stability of our apparatus, though, easily enables performing experiments lasting even several hours.

\subsection{\label{sec:Bi2212}Application on Y-Bi2212}
\begin{figure*}
    \includegraphics[scale=0.5]{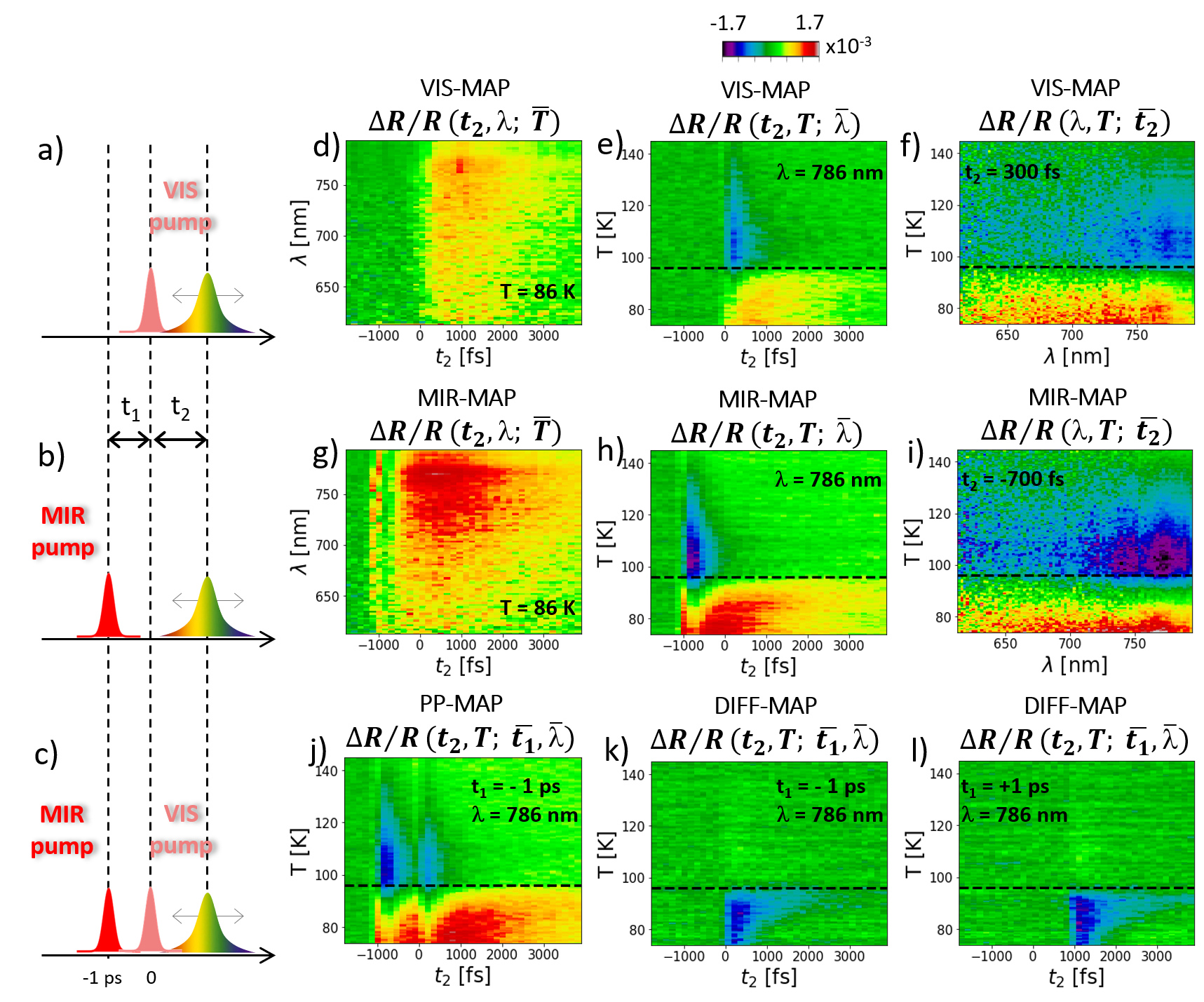}
    \caption{Transient reflectivity colour-coded maps. a), b) and c) show a scheme of the pulses involved and their temporal delays. Each scheme refers to the maps that lie in its row. VIS-MAPs (pump-probe signal triggered by the visible pump) are shown in d) as function of probe wavelength and temporal delay between the visible pump and the probe at fixed temperature T = 86 K; in e) as function of temperature and delay at fixed probe wavelength $\lambda$ = 786 nm; and in f) as function of temperature and probe wavelength at fixed positive delay t$_2$ = 300 fs. g), h) and i) follow the same structure as d), e) and f), but they now refer to the mid-IR pump-probe signal, as shown in b). Note that the dependence on (fixed) t$_1$ has been dropped in the titles of d-i) as it is not relevant for simple two-colour signals as VIS-MAPs and MIR-MAPs. j) is an example of PP-MAP showing the combined time-dependent signal at fixed wavelength $\lambda$ = 786 nm due to a mid-IR pump delayed by t$_1$ = -1 ps with respect to the visible one arriving at t$_2$ = 0 fs (as pictured in c)). k) and l) are DIFF-MAPs as function of temperature and delay at fixed wavelength and fixed delay t$_2$ between the two pumps: the mid-IR pumps arrives before and after the visible one by 1 ps in k) and l), respectively.}
    \label{fig:Bi2212}
\end{figure*}

To test the potential of our setup, we performed transient reflectivity measurements on a single crystal of Bi$_2$Sr$_2$Y$_{0.08}$Ca$_{0.92}$Cu$_2$O$_{8+\delta}$ (Y-Bi2212). The sample under exam is slightly underdoped and displays the superconducting transition at T$_C$ = 96 K. The crystal has been mounted so that the probe’s propagation axis is almost perpendicular to the ab-plane of the unit cell (i.e. the Cu-O planes). In our experiment, we first pump the sample using pulses centered at 1.44 eV (860 nm), and then push it with mid-IR excitations whose photon energy is closed to the superconducting gap of the sample (2$|\Delta| \simeq$ 75 meV). Both pumps’ polarizations have been chosen to be parallel to the [110] crystallographic direction (the Cu-Cu axis), while the probe is cross-polarized. We have worked using the full repetition rate of the laser (50 kHz) as we have witnessed no significant thermal effects in the sample. We have instead set the acquisition trigger to 2 kHz and used the NMOS array detectors. The pumps’ fluences absorbed by the sample are 40 $\mu$Jcm$^{-2}$ for the visible pump, and 100 $\mu$Jcm$^{-2}$ for the mid-IR one. In order to cover the relevant transitions in our samples, we ran a temperature scan starting from 74 K (superconducting phase) up to 144 K, which is reasonably close to the presumed transition temperature T* at our doping.

As detailed in the previous subsection, one single measurement is sufficient to obtain a great wealth of information. The results of the experiment on Y-Bi2212 are summarized in Figure \ref{fig:Bi2212}. In Figure \ref{fig:Bi2212}a, \ref{fig:Bi2212}b and \ref{fig:Bi2212}c, we provide simple schemes to clarify which pulses are involved in the colour-maps shown and give details on the time structure of the measurement. In this specific measurement, we chose to investigate just two time-delays between the pumps, t$_1=\pm$1 ps: when t$_1$=-1 ps, the mid-IR pump impinges on the sample 1 ps before the arrival of the visible one; conversely, when t$_1$=+1 ps, the mid-IR pump arrives 1 ps after. From the physical point of view, this corresponds to two different questions. In the former case, we are interested in the effect that low-photon energy excitations may have on the high-energy dynamics; in the latter case, the situation is reversed, and we investigate whether and how visible pulses can alter the dynamics triggered by low-energy excitations. It is thus clear that measurements of this kind could represent a valuable tool to study the intrinsic coupling between degrees of freedom at different energy scales, that lies at the very core of high temperature superconductivity. The detailed interpretation of this specific measurement and its importance to the physics of cuprates, goes beyond the scope of this paper and will be the subject of a future dedicated manuscript. Here we shall confine the discussion to the capabilities of the setup and to the type of information that can be extracted.

On the first row of Figure \ref{fig:Bi2212}, we display the VIS-MAPs, that is the transient reflectivity colour-coded maps that involve just the action of the visible pump. According to the scheme in Figure \ref{fig:data_structure}, we dissect the four-dimensional dataset into three different two-dimensional matrices and separately study the dependencies. In Figure \ref{fig:Bi2212}d, we show the wavelength-dependent pump-probe traces associated to the superconducting phase, which is consistent with ref. \cite{giannetti2011revealing}. A second map can be computed by fixing the wavelength and displaying the temperature dependence, as in Figure \ref{fig:Bi2212}e for $\lambda$ = 786 nm. The horizontal dashed black line highlights the sample’s critical temperature, marking the beginning of the pseudogap phase, which is accompanied by a change in both sign and dynamics of the transient reflectivity. Finally, we display in Figure \ref{fig:Bi2212}f $\Delta R/R$ as function of temperature and probe wavelength for a given time delay, showing that we do observe a wavelength-dependent response mainly in the pseudogap region of the phase diagram.

The above-discussed structure is the same for the MIR-MAPs in Figure \ref{fig:Bi2212}g, \ref{fig:Bi2212}h and \ref{fig:Bi2212}i, with the important difference that now the dynamics is triggered by mid-IR excitations. By comparing the VIS-MAPs and the MIR-MAPs, clear differences can be spotted that, at least in this measurement, could be attributed to the higher fluence of the mid-IR pump. Single-colour pump-probe traces (horizontal cuts in Figure \ref{fig:Bi2212}g) are indeed compatible with those observed in ref. \cite{giannetti2009discontinuity}. We highlight that Figure \ref{fig:Bi2212}i (and obviously its time dependence, namely cuts at different t$_2$) represents, to the best of our knowledge, the first mid-IR pump-broadband probe response ever measured.

In Figure \ref{fig:Bi2212}j, we show the PP-MAP which contains the interaction with both the visible and mid-IR pumps. Here we show just the two-dimensional cut for a given probe wavelength, as function of time-delay and temperature, but obviously a comprehensive analysis can be carried out according to the scheme in Figure \ref{fig:data_structure}. It is worth recalling that now a further degree of freedom becomes relevant, namely the relative time-delay of the two pumps, t$_1$. Figure \ref{fig:Bi2212}j displays just one possible choice of t$_1$, that is t$_1$ = -1 ps (as depicted in Figure \ref{fig:Bi2212}c); an additional, similar map must be considered for t$_1$ = +1 ps, where now the onset of the signal associated to the mid-IR pulse is shifted to t$_2$ = 1 ps.

Finally, Figure \ref{fig:Bi2212}k and \ref{fig:Bi2212}l show the DIFF-MAPs for t$_1$ = $\pm$1 ps, respectively. Again, we limit our discussion to two-dimensional slices at fixed wavelength only. Interestingly, the fact that these maps are not completely zero, is a clear indication that the combined action of the two pumps is not a mere superposition of the signals independently triggered by the visible and mid-IR pulses. In particular, in both cases, the difference is non-zero below the critical temperature, suggesting that the condensate is mostly affected by the subsequent photoexcitations. This could be a hint of an inherent coupling whose nature and dynamics will be further studied.

\section{Conclusions}
In this paper, we have reported the construction and commissioning of a novel optical setup for ultrafast broadband spectroscopy based on a three-pulse technique. The uniqueness of the experimental design lies in the simultaneous presence of two pumps of different tunable colour to selectively excite different degrees of freedom of the material, both in the visible and in the mid-infrared energy range. This configuration allows for the direct study and control of the coupling between low- and high-energy excitations in matter and addresses the interdependence of their dynamical response. The time evolution of the sample reflectivity (or transmittivity) is probed by a supercontinuum white-light probe, which allows to model the non-equilibrium dielectric function of the sample over a wide range of cryogenic temperatures. A dedicated acquisition electronics has been developed to enable referenced single-pulse detection up to 50 kHz, hence exploiting the full repetition rate of our laser. We tested the performance of our setup by performing transient reflectivity measurements on a single-crystal superconducting Y-Bi2212, a well-studied member of the cuprate family. We showed that, by running just a single experiment and thus suppressing artefacts, it is possible to single out the time delay-, probe wavelength-, and temperature-dependence of the non-equilibrium reflectivity. While we gave here just a glimpse of what kind of measurements and analysis can be carried out, the considerable size of the parameter space that can be explored by our setup paves the way to several and diverse applications, ranging from excited state absorption studies to optical control of organic photovoltaic materials, where, by coherently exciting low-energy vibrations, it could be possible to actively tune their charge transfer rate \cite{falke2014coherent,bakulin2015mode,desio2016tracking}, to the optical control of the superconducting transition through double quench techniques.

\section*{Acknowledgments}
We thank F. Boschini and A. Damascelli for providing the Y-Bi2212 sample. This work was supported by the European Research Council through the project COBRAS (ERC-2019-PoC, grant Agreement No. 860365) and the project INCEPT (ERC-2015-STG, grant Agreement No. 677488).

\bibliographystyle{ieeetr}
\bibliography{aabib.bib}

\end{multicols}

\end{document}